\documentclass[a4paper,fleqn,usenatbib]{mnras}

 \usepackage[normalem]{ulem}
 
\usepackage[T1]{fontenc}
\usepackage{ae,aecompl}

\usepackage{graphicx}	
\usepackage{amsmath}	
\usepackage{amssymb}	
\usepackage{multirow}
\usepackage{multicol}
\usepackage{longtable}
\usepackage{supertabular}
\usepackage{natbib}
\usepackage{hyperref}

\newcommand{\emm}[1]{\ensuremath{#1}} 
\newcommand{\emr}[1]{\emm{\mathrm{#1}}}
\newcommand{\chem}[1]{\ensuremath{\mathrm{#1}}} 
\newcommand{\unit}[1]{\emr{\,#1}} 
 
\newcommand{\MHz}{\unit{MHz}} 
\newcommand{\kHz}{\unit{kHz}} 
\newcommand{\radec}[6]{\emr{\alpha_{2000}=#1^{h}#2^{m}#3^{s}},
	\emr{\delta_{2000}=#4^{\circ}#5^{'}#6^{''}}} 
\newcommand{\Kkms}{\unit{K\,km\,s^{-1}}} 
\newcommand{\kms}{\unit{km\,s^{-1}}} 
\newcommand{\ps}{\unit{s^{-1}}} 
\newcommand{\K}{\unit{K}}
\newcommand{\sciexp}[2]{\emm{#1\times10^{#2}}}
\newcommand{\pscm}{\unit{cm^{-2}}}
\newcommand{\pccm}{\unit{cm^{-3}}}

\title[ CH$_3$SH in IRAS 16293-2422]{Detection of CH$_3$SH in protostar IRAS 16293-2422}

\author[Majumdar et al.]{
L. Majumdar$^{1,2,3}$\thanks{E-mail: liton.icsp@gmail.com}, P. Gratier$^{1,2}$, T. Vidal$^{1,2}$, V. Wakelam$^{1,2}$, J.-C. Loison$^{4,5}$,
\newauthor\  K. M. Hickson$^{4,5}$, E. Caux$^{6,7}$ 
\\ 
$^{1}$ Univ. Bordeaux, LAB, UMR 5804, F-33270, Floirac, France.\\
$^{2}$ CNRS, LAB, UMR 5804, F-33270, Floirac, France\\
$^{3}$ Indian Centre For Space Physics, 43 Chalantika, Garia Station Road, Kolkata, 700084, India\\
$^{4}$ Univ. Bordeaux, ISM, UMR 5255, F-33400 Talence, France\\
$^5$ CNRS, ISM, UMR 5255, F-33400 Talence, France\\
$^{6}$Universit\'e de Toulouse, UPS-OMP, IRAP, Toulouse, France\\
$^{7}$CNRS, IRAP, 9 Av. Colonel Roche, BP 44346, F-31028 Toulouse Cedex 4, France
}

\date{Accepted XXX. Received YYY; in original form ZZZ}

\pubyear{2015}

\begin{document}
\maketitle

\begin{abstract}
The nature of the main sulphur reservoir in star forming regions is a long standing mystery. The observed abundance of sulphur-bearing species
in dense clouds is only about 0.1 per cent of the same quantity in diffuse clouds. Therefore, the main sulphur species in star forming regions 
of the interstellar medium are still unknown. IRAS 16293-2422 is one of the regions where production of S-bearing species is favourable due to 
its conditions which allows the evaporation of ice mantles.
We carried out observations in the 3 mm band towards the solar type protostar IRAS 16293-2422 with the IRAM 30m telescope.
We observed a single frequency setup with the EMIR heterodyne 3 mm receiver with an Lower Inner (LI) tuning frequency of 89.98 GHz. 
Several lines of the complex sulphur species CH$_3$SH were detected. Observed abundances are compared with simulations using the NAUTILUS 
gas-grain chemical model. Modelling results suggest that CH$_3$SH has the constant abundance of $4 \times10^{-9}$ (compared to H$_2$) for radii lower than 200 AU and is mostly 
formed on the surfaces.
 Detection of CH$_3$SH indicates that there may be several new 
families of S-bearing molecules (which could form starting from CH$_3$SH) which have not been detected or looked for yet.
\end{abstract}




\section{Introduction}
Though sulphur is only the tenth most abundant element  in the interstellar medium (ISM) and stars, it has long been of interest for 
astrochemists. First of all, the reservoir of sulphur in dense regions is a matter of debate. The depletion of sulphur in the diffuse medium does not seem to occur as its atomic abundance in the gas-phase is rather constant with the line of sight \citep{1994ApJ...430..650S} even though these measurements are uncertain \citep{2009ApJ...700.1299J}. The observed abundance of sulphur-bearing species in dense interstellar media is still only a few percent of the cosmic reference  \citep{1994A&A...289..579T,2004A&A...413..609W,2013ApJ...779..141A}. The most simple explanation is that atomic sulphur depletes on interstellar grains with increasing density and it is converted into H$_2$S on the surface \citep{1997ApJ...481..396C}. However, the lack of a solid H$_2$S feature in the original signature in ISO spectra of high \citep{2000ApJ...536..347G} and low \citep{2000A&A...360..683B} mass protostars sets an upper limit on the mantle H$_2$S abundance which cannot exceed about 10$^{-7}$ with respect to H$_2$ \citep{1998ARA&A..36..317V}. Other more or less refractory (polymers or aggregates of sulphur) have then been proposed as reservoirs for interstellar sulphur \citep{2004A&A...422..159W,2012MNRAS.426..354D,2015MNRAS.450.1256W}. Whatever its form, it is now acknowledged that sulphur is mostly present on grain surfaces in dense environment. In the envelopes of protostars, through thermal desorption, this sulphur returns to the gas-phase and produces a chemical chain forming first SO and then SO$_2$ \citep{1997ApJ...481..396C,2003A&A...399..567B,2004A&A...422..159W}. Assuming that part of the initial sulphur is in the form of H$_2$S and/or OCS, the relative abundance ratios of these four species could be in principle used as chemical clocks of high and low mass protostars \citep{1998A&A...338..713H,2011A&A...529A.112W,2015ApJ...802...40L}.

CS was the first sulphur bearing molecule detected in the interstellar medium \citep{1971ApJ...168L..53P}. Since then, several sulphur-bearing molecules have been detected in the interstellar medium and circumstellar shells in the Milky Way \citep[see for instance ][and references therein]{2015MNRAS.446.3118B}.  The most complex S-bearing molecule, ethyl mercaptan CH$_3$CH$_2$SH, was recently detected by \citet{2014ApJ...784L...7K} in Orion. Previously methyl meracaptan CH$_3$SH was first detected in SgrB2 \citep{1979ApJ...234L.139L} followed by observations in the organic-rich hot core G327.3-0.6 \citep{2000ApJ...545..309G}, and in Orion \citep{2014ApJ...784L...7K}. \citet{2000ApJ...545..309G} suggested that CH$_3$SH  forms 
in the ices and then evaporates in hot cores. This paper reports the detection of CH$_3$SH in IRAS 16293-2422 followed by astrochemical modelling in order to explain the observed abundances. The observations and analysis are presented in Section 2. The chemical model is described in Section 3 while the results are discussed in the last Section.
\section{Observations and data reduction}

\subsection{Observations}
Observations were carried out using the IRAM 30m telescope from the 18th to the 23rd of August in average summer conditions (a median value of 4-6 mm water vapor).
We used the EMIR heterodyne 3 mm receiver tuned at a frequency of 89.98 GHz
in the Lower Inner sideband. The receiver was followed by a Fourier Transform Spectrometer
in its 195\kHz{} resolution mode, the observed spectrum is composed of two
approximately 8 GHz regions centered respectively on 88.41 and 104.06 GHz.

The beam size corresponding to the median value of the frequencies of the observed lines is $24.5''$ and this corresponds to 2940 AU 
for a distance of 120 pc (Crimier et. al 2010). The separation between source A and B is $5.5''$, so that both sources are well within the telescope beam. 
The position \radec{16}{32}{22.75}{-24}{28}{34.2}, midway between sources A
and B of IRAS16293 was observed using the wobbler switching mode with a
period of 2 seconds and a throw of $90''$ ensuring mostly flat baselines
even in summer conditions and observations at low elevation. The nearby
planet Saturn was used for focus (at the beginning of each run and after
sunset) and pointing (every hour) with mostly good pointing corrections
(less than a third of the beam size).

 \subsection{Results}
 \begin{table*}
    \caption{\label{tab.obs}Spectroscopic data for the observed lines of \chem{CH_3SH} and derived line parameters.}
    \centering
    \begin{tabular}{llllllll}
        \hline
        Lines& W           & V$_{\rm LSR}$          & FWHM & Frequency & Aij & Eup  & QNs\\
                & (\Kkms)        & (\kms)       & (\kms) & (\MHz) & (\ps) & (\K) & \\
        \hline
        1                  & $59.31\pm5.01$                  & $3.6\pm0.2$                  & $4.2\pm0.4$                  & 101029.81 & \sciexp{2.15}{-6} & 11.8  & $4_{1,4} \rightarrow 3_{1,3}$ (E$-$)\\
        \hline
        \multirow{2}{*}{2} & \multirow{2}{*}{$122.05\pm4.4$} & \multirow{2}{*}{$4.4\pm0.7$} & \multirow{2}{*}{$3.8\pm0.2$} & 101139.16 & \sciexp{2.31}{-6} & 7.3   & $4_{0,4} \rightarrow 3_{0,3}$ (A$+$)\\
                           &                                 &                              &                              & 101139.65 & \sciexp{2.31}{-6} & 8.7   & $4_{0,4} \rightarrow 3_{0,3}$ (E$+$)\\
        \hline
        \multirow{4}{*}{3} & \multirow{4}{*}{$49.01\pm4.3$}  & \multirow{4}{*}{$4.5\pm0.3$} & \multirow{4}{*}{$6.9\pm0.6$} & 101159.46 & \sciexp{1.73}{-6} & 26.4  & $4_{2,3} \rightarrow 3_{2,2}$ (A$-$)\\
                           &                                 &                              &                              & 101160.53 & \sciexp{1.01}{-6} & 47.5  & $4_{3,2} \rightarrow 3_{3,1}$ (E$-$)\\
                           &                                 &                              &                              & 101160.53 & \sciexp{1.01}{-6} & 47.7  & $4_{3,2} \rightarrow 3_{3,1}$ (A$+$)\\
                           &                                 &                              &                              & 101160.53 & \sciexp{1.01}{-6} & 47.7  & $4_{3,1} \rightarrow 3_{3,0}$ (A$-$)\\
        \hline
        \multirow{2}{*}{4} & \multirow{2}{*}{$53.70\pm5.5$}  & \multirow{2}{*}{$3.6\pm0.3$} & \multirow{2}{*}{$6.3\pm0.8$} & 101167.15 & \sciexp{1.73}{-6} & 24.8  & $4_{2,3} \rightarrow 3_{2,2}$ (E$-$)\\
                           &                                 &                              &                              & 101168.34 & \sciexp{1.73}{-6} & 25.4  & $4_{2,2} \rightarrow 3_{2,1}$ (E$+$)\\
        \hline
        5                  & $40.90\pm5.03$                  & $3.7\pm0.2$                  & $2.9\pm0.4$                  & 101284.36 & \sciexp{2.17}{-6} & 13.5  & $4_{1,3} \rightarrow 3_{1,2}$ (E$+$)\\
        \hline
        6                  & $51.51\pm3.62$                  & $3.7\pm0.2$                  & $4.4\pm0.4$                  & 102202.43 & \sciexp{2.23}{-6} & 12.4  & $4_{1,3} \rightarrow 3_{1,2}$ (A$-$)\\

        
        \hline
    \end{tabular}
\end{table*} 

\subsubsection{\chem{CH_3SH} line properties}

\begin{figure*}
    \includegraphics[width=\textwidth]{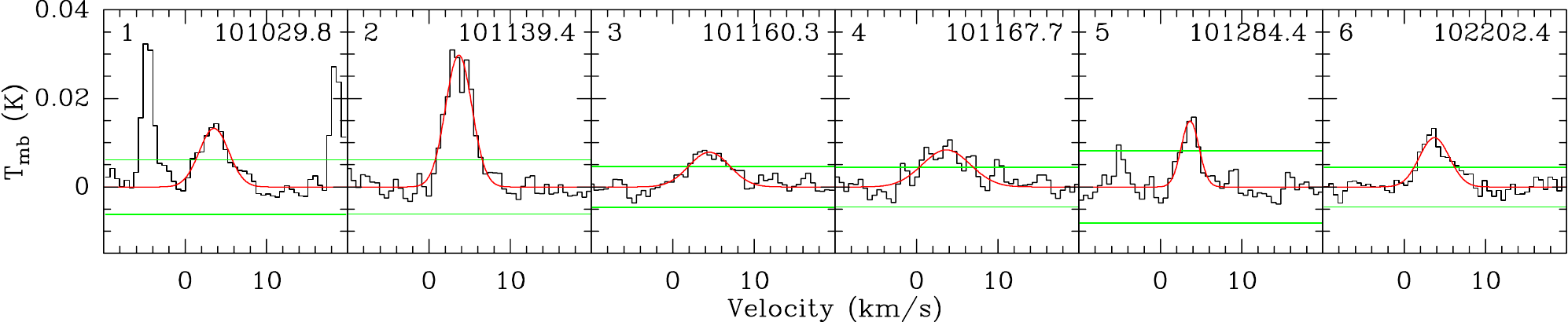}
    
    \caption{\label{fig.gauss_fit} Gaussian fits to the observed line
    attributed to \chem{CH_3SH}. The number in the top left of each panel
    corresponds to the line number in Table~\ref{tab.obs}. The two additional
    lines in the first panel are \chem{H_2CCO} lines. The green lines are $\pm 3\sigma$ levels. }
\end{figure*}

We used the CLASS software from the
GILDAS\footnote{\url{https://www.iram.fr/IRAMFR/GILDAS/}} package to reduce
and analyse the data. Gaussian fits were made to the detected lines
following a local low (0 or 1) order polynomial baseline subtraction.
Table~\ref{tab.obs} and Fig.~\ref{fig.gauss_fit} show the result of these fits
for the 6 observed lines of \chem{CH_3SH}. 
Some observed lines correspond to unresolved spectral components, in these cases a single gaussian was fitted, which 
is often wider than the line corresponding to a single component, this is particularly important for lines 3 and 4, having spectral components separated by $\sim$ 3 km/s. For the 3 single component features, the mean LSR velocity is 3.67\kms{} and the
mean FWHM is 3.8\kms{}. \citet{2011A&A...532A..23C} established a classification of the
species in a FWHM vs VLSR plane, \chem{CH_3SH} is similar to the cluster
identified as Type IV not associated specifically to any of the two spatial
components A or B of IRAS16293, these species either come from a common envelope or are
found in the two components.

\subsubsection{\chem{CH_3SH} radiative transfer modelling}

We use an LTE approach to model the emission of \chem{CH_3SH}, the
parameters in the radiative transfer model are the species column density,
the line width, the excitation temperature and the source size (we assume an
axisymmetric gaussian source centred in an axisymmetric gaussian beam whose
size is dependent on the observed frequency). The model also requires 
accurate molecular spectroscopic data in the form of a catalog file which
contains energy levels with associated quantum numbers, statistical weights
and transition frequencies as well as the integrated intensity at 300 K.
For \chem{CH_3SH}, it was retrieved from the CDMS
\citep{2005JMoSt.742..215M} using the VAMDC portal
\citep{2010JQSRT.111.2151D} and the frequencies, Einstein coefficients and
upper level energies are listed in Table~\ref{tab.obs}. When several
lines of \chem{CH_3SH} are blended in a single observed profile, we sum the
corresponding modelled integrated intensities. This is appropriate in the low
opacity regime observed for \chem{CH_3SH} in IRAS16293.

We use a bayesian approach to recover the distribution of parameters which
best agree with the observed line intensities. The likelihood is constructed
assuming gaussian centred noise with a magnitude defined by the observed
uncertainties on the observed integrated areas. The prior are chosen to be
uniform and non informative over the range of variation as defined in
Table~\ref{tab.priors}, except for the line width for which a gaussian prior
of mean 3.66 and standard deviation 0.2, corresponding to the median and
typical errors of the single component gaussian line fits. This is done
because in the optically thin regime where \chem{CH_3SH} emits in IRAS16293,
the integrated intensity is independent of the line width.

The posterior distribution function is sampled using the Affine Invariant
Ensemble Monte Carlo Markov Chain approach described in \citet{Goodman2010}
in its Python implementation {\tt
EMCEE}\footnote{\url{https://github.com/dfm/emcee}}
\citep{2013PASP..125..306F}. One hundred walkers are initialised uniformly
within the parameter ranges used and the chains are evolved for a burning
sequence of 20000 steps after which the convergence is checked by examining a
plot of the running mean of each parameters.

\begin{figure}
    \includegraphics[width=\columnwidth]{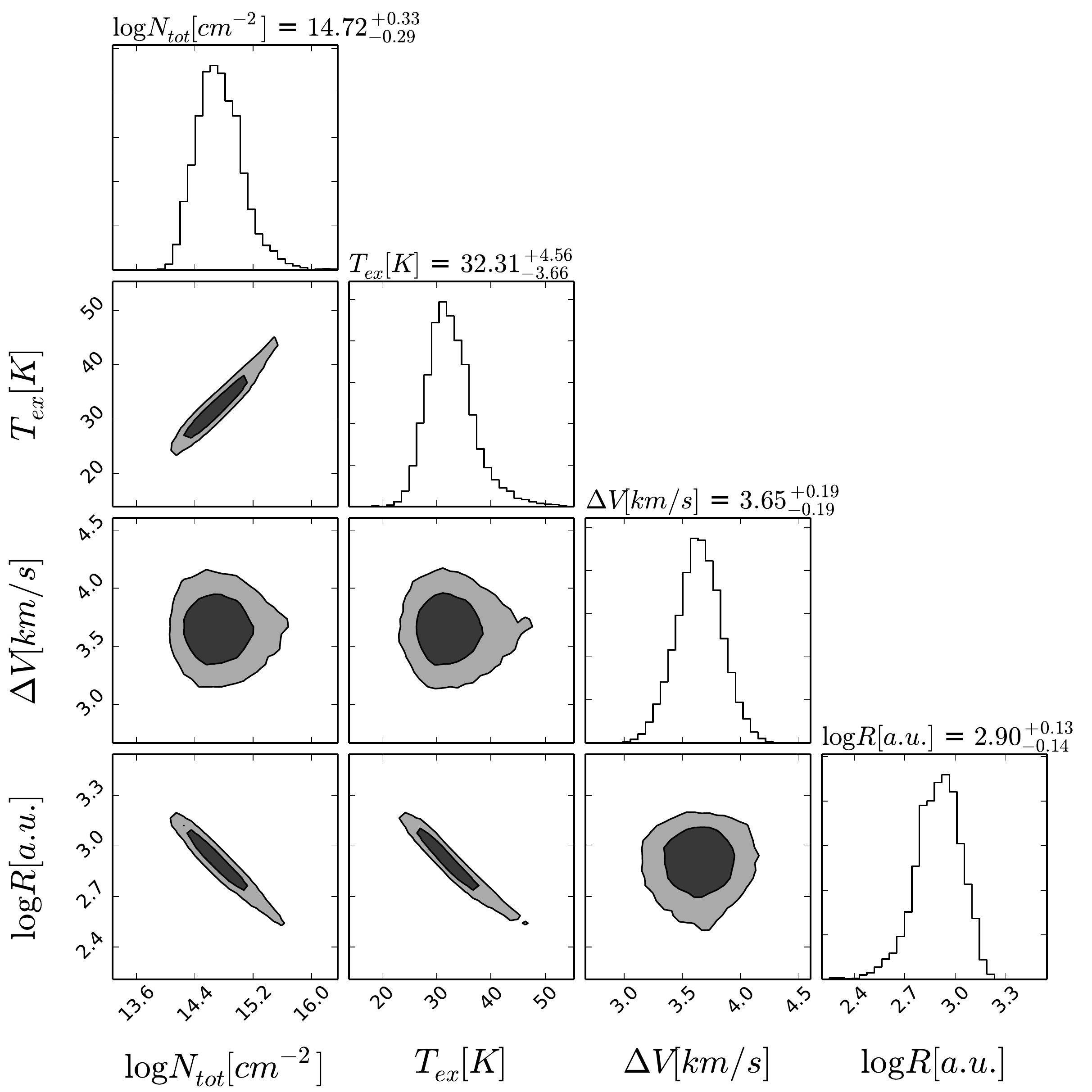}
    \includegraphics[width=\columnwidth]{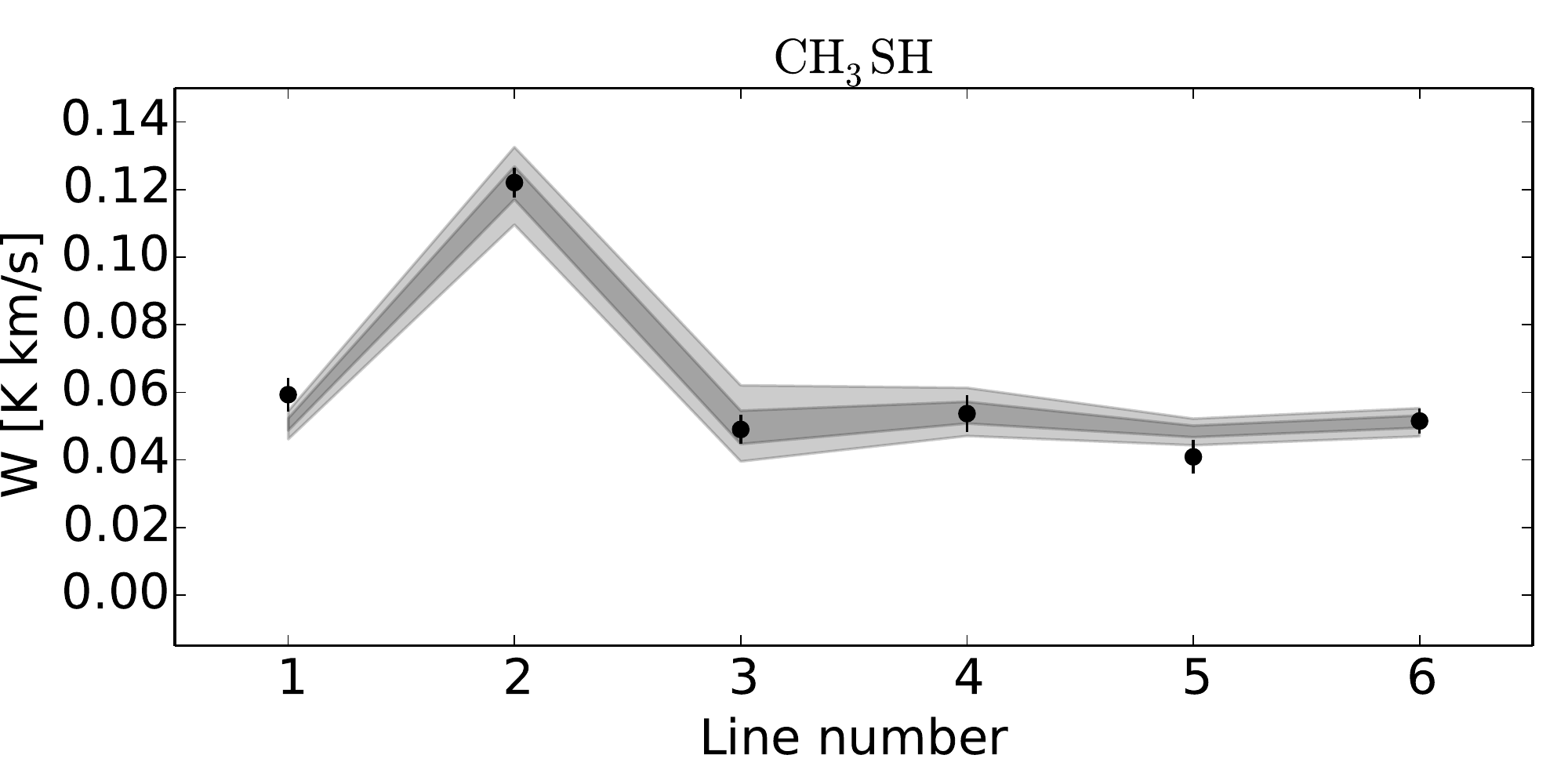}
    \caption{\label{fig.bayes_result} 1D and 2D histograms of the
   posterior distribution of parameters. Contours contain respectively 68
   and 95 percent of samples. Uncertainties are at the 16th and 84th
   percentile around the median value of the distribution. Bottom: observed
   integrated intensities and 1$\sigma$ and 2$\sigma$ distribution of the models
   corresponding to the posterior distribution of parameters. Lower axis
   indices correspond to the line numbers in Table 1.}
    \end{figure}

Figure~\ref{fig.bayes_result} shows the 1D and 2D histogram of the posterior
probability distribution function and the comparison of the observations
with the distribution of computed intensities corresponding to the posterior
distribution of parameters. Table~\ref{tab.bayes_result} summarises the
point estimates for the parameters. The \chem{H_2} column density is
computed by integrating the power law in \citet{2010A&A...519A..65C} up to
the size derived by the bayesian method. The \chem{H_2} column density
corresponding to the derived distribution of radius is \sciexp{1.11 \pm
0.02}{24}\pscm{}, the uncertainties on the abundance will be dominated by
the uncertainties on the \chem{CH_3SH} column density. The distribution of line opacities of the 
observed line can also be determined from the posterior distribution of the parameters. The derived
line opacities range from $\sciexp{4}{-3}$ to $\sciexp{2}{-2}$ validating the optically thin
hypothesis necessary to sum the blended line intensities.

It is not possible to infer the source size solely from the observed line intensities. As a matter of fact, we have 
observed the source in a limited range frequency and as a consequence, the different beam sizes are very close 
each other (only $~1\%$ relative variation). However, the range in upper level energies is large enough that the
excitation temperature can be estimated accurately. By assuming LTE
conditions, we have an estimate of the gas temperature of the region emitting
in \chem{CH_3SH}. \citet{2010A&A...519A..65C} have determined the temperature
and density profile of IRAS16293 as a function of radius. By using a
parametrization of the gas temperature as a function of radius based on
their Fig. 7, it is possible to attribute a source size for any given gas
temperature. This estimated size can then be used to derive the overall beam
dilution in the telescope beam and compute a column density corrected from
the beam dilution. This reasoning is applied in a self coherent manner to
determine all the model parameters from the dataset.

There remains the possibility that emission of \chem{CH_3SH} coming from hotter
gas in the more central parts of the pre-stellar envelope could also be
detected and bias the measured parameters. It is possible to check that this
contamination is negligible in our case. Indeed, emission from the inner
hotter parts of the pre-stellar envelope at a temperature of 200 K (6 times
larger than the 30 K computed) is associated with a source size of only 10 AU
(80 times smaller than the 760 AU computed). \chem{CH_3SH} emission from the
hotter central parts will be $80^2/6 \sim1000$ times less bright than the
detected emission. This is much lower than the detection limit. In terms of abundances 
towards the inner part, this translates into very high upper limits $[\chem{CH_3SH}] < \sciexp{5}{-8}$, the chemical modelling
results showing a constant abundance of $4\times 10^{-9}$ for radii lower
than 200 AU is thus compatible with the observations.

\begin{table}
   \caption{Priors distribution functions for the parameters used in the
   bayesian approach.}
   \label{tab.priors}
   \centering
   \begin{tabular}{ll}
       \hline
       Parameter & Distribution  \\
       \hline
                    \emr{\log N} &              $\emr{Uniform}(9,22 )$ \\
                    \emr{T_{ex}} &              $\emr{Uniform}(3,200)$ \\
                    \emr{\Delta V} &            $\emr{Normal}(3.66,0.2)$\\
       \hline
   \end{tabular}
   \begin{minipage}{8cm}Notes:
   \emr{ Uniform}(min-value, max-value) is a uniform distribution with
   values going from minimum value to maximum value, Normal($\mu$,
   $\sigma$) is a gaussian distribution with mean $\mu$ and standard
   deviation $\sigma$ \end{minipage}
\end{table}

%
%

\begin{table}
    
    \caption{Point estimates of the posterior distribution function
    corresponding to the median and one sigma uncertainty.}
    
    \label{tab.bayes_result}
    \centering
    \begin{tabular}{lc}
        \hline
        Parameter           & Value  \\
        \hline
           \emr{\log N (CH_3SH)} (\pscm)   &   $14.7\pm0.3$         \\
           \emr{T_{ex}} (\K)      &   $32    \pm4$         \\
           \emr{\Delta V} (\kms)  &   $3.7\pm0.2$         \\
           \emr{\log R} (a.u.)    &   $2.90\pm0.13$         \\
           \emr{\log [CH_3SH]^{a}}    &   $-9.3\pm0.3$         \\
        \hline
        \multicolumn{2}{l}{Notes: \emr{^{a}} [X] = N(X)/N(\chem{H_2})}
    \end{tabular}
\end{table}

\subsubsection{Validity of LTE}

The LTE approach is only valid if the number density is much larger than the
critical density of a given molecular transition. Using the density profile derived
by \citet{2010A&A...519A..65C}, we find that the density of \chem{H_2} is of the
order of \sciexp{3}{6}\pccm{}. The collisional (de)excitation rates are not
known for \chem{CH_3SH}, however we can approximate their value to the ones
of methanol or methylcyanide for similar transitions: $\sim 10^{-11}
\unit{cm^{3}\ps}$, with typical Einstein coefficients of $\sim
10^{-6}\ps$, this yields a critical density of $\sim 10^{5}\pccm$, more than
an order of magnitude below the \chem{H_2} density. Since the density of IRAS16293 is relatively high, we expect that the abundances 
derived in the LTE approximation are only moderately underestimated. The similar type of approach was considered in the past for analysing Complex Organic 
Molecules (COMs) in IRAS16293 \citep{2014ApJ...791...29J}.

\section{Chemical modeling \& network}

\subsection {The NAUTILUS chemical model}
To simulate the abundance of CH$_3$SH in IRAS16293, the NAUTILUS gas-grain chemical model \citep{2010A&A...522A..42S,2014MNRAS.440.3557R} has been used with spherical protostellar core physical conditions similar to \citet{2008ApJ...674..984A,2012ApJ...760...40A}. NAUTILUS is a state of the art chemical code which computes the abundances of species (e.g. atoms, ions, radicals, molecules) as a function of time in the gas phase and at the surface of the interstellar grains. All the equations and physicochemical  processes included in the model were discussed in detail in \citet{2014MNRAS.440.3557R} and in \citet{2015MNRAS.447.4004R}. 

By following the kida.uva.2014 chemical network of \citet{2015ApJS..217...20W}, several types of chemical reactions are considered in the gas phase. For example: bimolecular reactions between neutral species, between charged species and between neutral and charged species, unimolecular reactions  i.e. photoreactions with direct UV photons and UV photons produced by the deexcitation of H$_2$ excited by cosmic ray particles, and direct ionisation and dissociation by cosmic ray particles. The grain surface is modeled by a one phase rate equation approach \citep{1992ApJS...82..167H}, i.e. there is no differentiation between the species in the bulk and at the surface. The interactions of gas-phase species with the interstellar grains are considered via four major steps: physisorption of gas phase species onto grain surfaces, diffusion of the accreted species, reaction at the grain surface and finally by evaporation to the gas phase. Our model also considers different types of evaporation process such as thermal evaporation, evaporation induced by cosmic rays \citep[following][]{1993MNRAS.261...83H}, and chemical desorption as suggested by \citet{2007A&A...467.1103G}. To simulate the chemistry of IRAS16293, we adopt the similar initial abundances reported  in \citet{2011A&A...530A..61H}, with an additional elemental abundance of $6.68\times 10^{-9}$ (compared to the total proton density) for fluorine \citep{2005ApJ...628..260N} and a C/O elemental ratio of 0.7 (i.e. the oxygen elemental abundance is $2.4\times 10^{-4}$). 

\subsection{Extension of the network}
To follow the chemistry of CH$_3$SH, we have added several new reactions that are listed in Table A1. 
Most of the reactions in Table A1 are defined based on similarities with methanol and by following the well known reactivity of carbon atoms. For the reactions at the surface of the grains, we assume that CH$_3$SH forms through successive hydrogenation reactions of CS similarly to CH$_3$OH since quantum chemical calculations show that all the steps are exothermic in nature. Among all these added surface reactions, there are barriers for s-\chem{H} + s-\chem{CS} and s-\chem{H }+ s-\chem{H_2CS} reactions but smaller than for s-\chem{H} + s-\chem{CO} and s-\chem{H}+ s-\chem{H_2CO} reactions. This can be explained by the d-orbitals of sulphur. As for both oxygen and sulphur, there is no occupied d orbital in the ground state so both 3d orbitals are vacant. But for sulphur, the energy needed to promote an electron from  3s or 3p orbital to 3d orbital is much less than the energy needed for an electron in the 2s or 2p orbitals in oxygen to 3d orbital. For this reason, the barrier we assumed for s-\chem{H }+ s-\chem{H_2CS} reaction is 1500 K. This value is close to the average of the calculated value at M06-2X/cc-pVTZ level (750 K) and MP2/cc-pVTZ level (2250K).
\subsection{The Protostellar physical model}
The chemistry of CH$_3$SH in IRAS 16293 is modeled using the same physical structure as in \citet{2008ApJ...674..984A} and \citet{2014MNRAS.445.2854W}, and was computed using the radiation hydrodynamical (RHD) model from \citet{2000ApJ...531..350M}. This model initially starts from a dense molecular cloud core with a central density $n$(H$_2$) $\sim 3 \times 10^4$ cm$^{-3}$ and the core is
 extended up to  $r=4 \times 10^4$ AU with a total mass of 3.852 $M_{\odot}$. The prestellar core evolves to the protostellar core in $2.5 \times 10^5$ yr. When the protostar is formed, the model again follows the evolution for $9.3 \times 10^4$ yr, during which the protostar grows by mass accretion from
the envelope. \citet{2010A&A...519A..65C} constrained the physical structure of the IRAS 16293 envelope through multi-wavelength dust and molecular observations.  The physical structure of the envelope used for this study at the final time of the simulation is similar to \citet{2010A&A...519A..65C}.

\begin{figure*}
  \includegraphics[width=\textwidth]{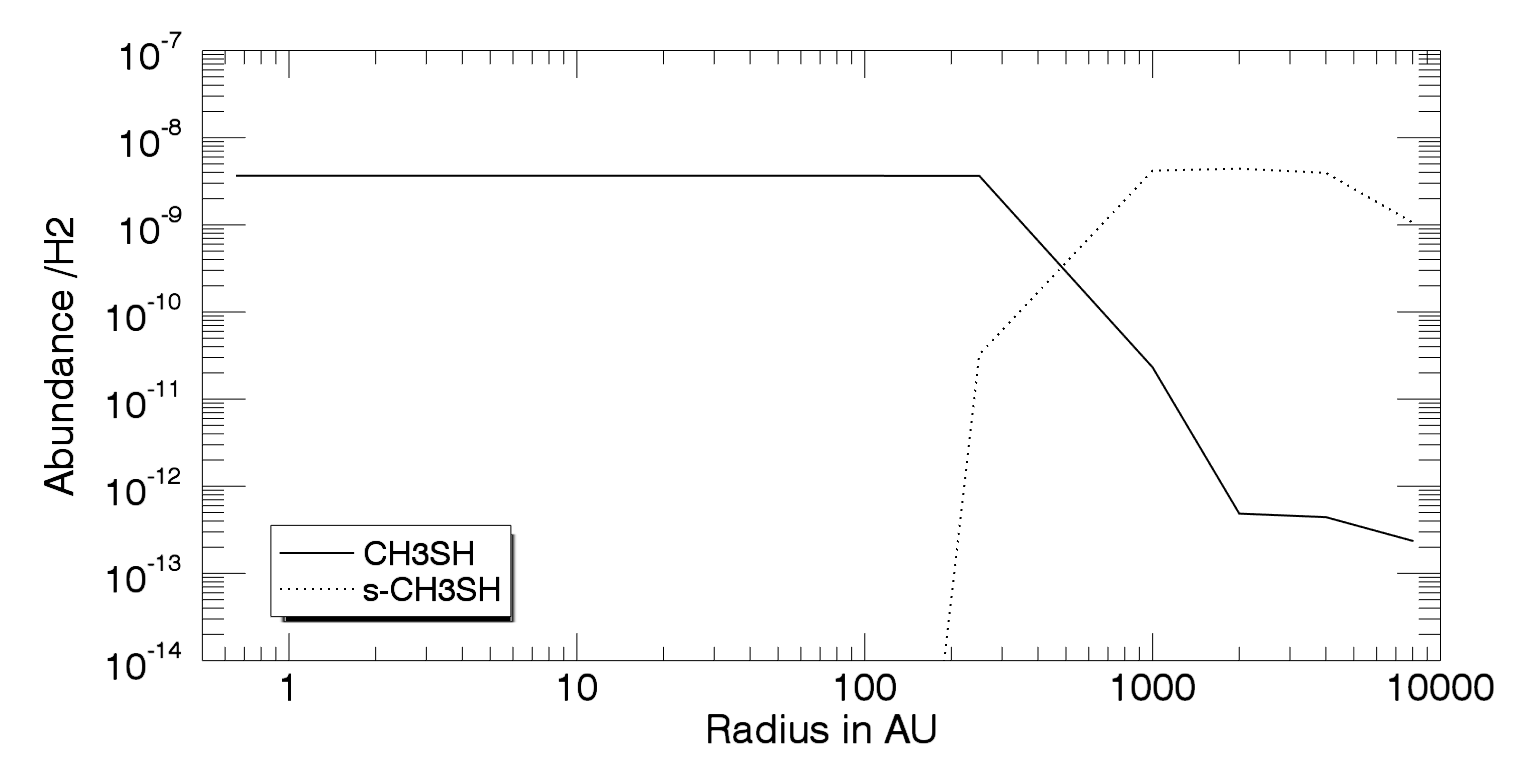}
  \caption{CH$_3$SH abundance (with respect to H$_2$) predicted by our model as a
function of radius. s-CH$_3$SH represents the CH$_3$SH on the surface of grains.}   
\end{figure*}

\section{Modeling Results and Discussions}
Figure 3 shows the computed abundances of CH$_3$SH, in the gas-phase and at the surface of the grains in the protostellar envelope as a function of radius to the central protostar. When the temperature increases above 30 K around 1000 AU, the gas-phase abundance of CH$_3$SH increases sharply up to $ 4 \times10^{-9}$. The CH$_3$SH abundance on the grains shows an inverse profile showing that at low temperature, the CH$_3$SH molecule is formed on the grains (in the outer part of the envelope) and is thermally desorbed in the inner part of the envelope when the temperature exceeds the evaporation temperature of the species. 
Here our model results have been obtained assuming a depleted sulphur elemental abundance of $8\times 10^{-8}$ (compared to H$_2$) for both the pre-collapse and the protostellar phases. We also ran our model assuming that the missing sulphur is released into the gas-phase in atomic form in the hot corino with an abundance of $1\times 10^{-5}$ (compared to H$_2$). The gas-phase abundance of CH$_3$SH is $5\times 10^{-8}$ (compared to H$_2$) in that case. The CH$_3$SH abundance predicted by our model in the gas phase is thus compatible with the observations. Another parameter that may influence the CH$_3$SH gas-phase profile in the protostellar envelope is the assumed binding energy, which is not known. In our model, we assumed a binding energy of 2700 K, which results in an evaporation of this species around 500 AU from the central star, corresponding to a temperature around 50 K. 

The identification of CH$_3$SH in a low mass protostar opens new opportunities to look for other complex sulphur bearing species. From Table A1, it is clear that H$_2$CS, CH$_3$S are some of the species related to the formation of CH$_3$SH. According to \citet{doi:10.1021/jp103357z}, reaction between CH$_3$SH and CH$_3$S, H$_2$CS can lead to (CH$_3$S)$_2$. \citet{doi:10.1021/jp0673415} has also shown that reaction between CH$_3$SH and OH can lead to CH$_3$SOH. Spectroscopic data for these species are however missing for the moment. 


\section*{Acknowledgements}

Based on observations carried out with the IRAM
30m Telescope. IRAM is supported by INSU/CNRS (France), MPG (Germany) and
IGN (Spain). LM, PG, VW and TV thanks ERC starting grant (3DICE, grant agreement 336474) for funding during this 
work. PG postdoctoral position is funded by the INSU/CNRS. VW, JCL and KMH acknowledge the CNRS programme PCMI for funding of their research. We  also  thank  the  anonymous  reviewer  for  his  
useful  comments  to  improve  the manuscript.






\bibliographystyle{mnras}
\bibliography{CH3SH}


\appendix

\section{List of added reactions}

\begin{table*}
\scriptsize{
\label{list_reactions}
\caption{List of gas-phase and grain surface reactions added to the model and associated parameters.}
\begin{center}
\begin{tabular}{|clc|c|c|c|c|c|c|c|}
\hline
\hline
& Reaction  & & $\alpha$ & $\beta$ & $\gamma$ & Reference\\
\hline
1 & H + CH$_3$S & $\rightarrow$ H$_2$CS + H$_2$ & $3.00\times 10^{-11}$ & 0 & 0 & 2\\
& & $\rightarrow$CH$_3$ + HS & $3.00\times 10^{-12}$ & 0 & 0 & 2 \\
\hline
2 & S + CH$_3$S & $\rightarrow$ H$_2$CS + HS & $6.00\times 10^{-12}$ & 0 & 0 & 9  \\
&& $\rightarrow$ CH$_3$ + S$_2$ & $1.90\times 10^{-11}$ & 0 & 0 & 9\\
\hline
3 &  C + CH$_3$S & $\rightarrow$ CH$_3$ + CS &$3.00\times 10^{-10}$ & 0 & 0 & 3 \\
\hline
4 & N + CH$_3$S & $\rightarrow$ H$_2$CS + NH & $1.00\times 10^{-11}$ &-0.17 & 0 & 9  \\
&&  $\rightarrow$ CH$_3$ + NS & $6.00\times 10^{-11}$ &0 & 0 & 9 \\
\hline
5 & O + CH$_3$S & $\rightarrow$ CH$_3$ + SO & $4.00\times 10^{-11}$ &0 & 0 &  5 \\
\hline
6 & C + CH$_3$SH & $\rightarrow$ CH$_3$ + HCS & $3.00\times 10^{-10}$ &0 & 0 & 4 \\
\hline
7 & CN + CH$_3$SH & $\rightarrow$ CH$_3$S + HCN & $2.70\times 10^{-10}$ &0 & 0 &  7 \\
\hline
8 & OH + CH$_3$SH & $\rightarrow$ CH$_3$S + H$_2$O & $8.00\times 10^{-12}$ &-0.4 & 0 & 6 \\
\hline
9 & H$_2$S + CH$_3$$^+$ & $\rightarrow$ CH$_3$SH$_2$$^+$ + Photon & $2.00\times 10^{-9}$ &0 & 0 & 10 \\
\hline
10 & CH$_3$SH + He$^+$ & $\rightarrow$ CH$_3$$^+$ + He + HS & $ 0.5 $ &  $2.82\times 10^{-9}$ & 3.32 & 1 \\
\hline
11 & CH$_3$SH + H$_3$$^+$ & $\rightarrow$ CH$_3$SH$_2$$^+$ + H$_2$ & $  3.20\times 10^{-9}$    &  2.3 & 3 & 1 \\
\hline
12 & CH$_3$SH + CH$_4$$^+$ & $\rightarrow$ CH$_3$SH$_2$$^+$ + CH$_3$ &$ 0.40 $ &  $1.56\times 10^{-9}$ & 3.32 & 1 \\
&& $\rightarrow$ CH$_3$SH$^+$ + CH$_4$ & $ 0.6 $ &  $1.56\times 10^{-9}$ & 3.32 &  1 \\
\hline
13 & CH$_3$SH + C$^+$ & $\rightarrow$ CH$_3$$^+$ + HCS & $ 0.80 $ &  $1.75\times 10^{-9}$ & 2.33 & 1 \\
&&  $\rightarrow$ CH$_3$S$^+$ + CH & $ 0.20 $ &  $1.75\times 10^{-9}$ & 2.33 &  1 \\
\hline
14 & CH$_3$SH + HCO$^+$ & $\rightarrow$ CH$_3$SH$_2$$^+$ + CO & $ 1.30\times 10^{-9} $ &  2.3 & 3 &  1 \\
\hline
15 & CH$_3$SH + N$^+$ & $\rightarrow$ CH$_3$ + H + NS$^+$ & $ 0.1 $ &  $1.64\times 10^{-9}$ & 3.32 &  1 \\
&& $\rightarrow$ CH$_3$$^+$ + H + NS & $ 0.04 $ &  $1.64\times 10^{-9}$ & 3.32 &  1 \\
&& $\rightarrow$ H$_2$CS$^+$ + H + NH & $ 0.3 $ &  $1.64\times 10^{-9}$ & 3.32 & 1 \\
& & $\rightarrow$ CH$_3$SH$^+$ + N & $ 0.4 $ &  $1.64\times 10^{-9}$ & 3.32 &  1 \\
\hline
16 & CH$_3$SH + O$^+$ & $\rightarrow$ H$_2$CS$^+$ + H$_2$O & $ 0.05 $ &  $1.56\times 10^{-9}$ & 3.32 &  1 \\
&& $\rightarrow$ CH$_3$SH$^+$ + O & $ 0.25 $ &  $1.56\times 10^{-9}$ & 3.32 &  1 \\
\hline
17 & CH$_3$SH + H$_3$O$^+$ & $\rightarrow$ CH$_3$SH$_2$$^+$ + H$_2$O & $ 1.50\times 10^{-9} $ &  2.3 & 3 &  1 \\
\hline
18 & CH$_3$SH + CH$^+$ & $\rightarrow$ CH$_3$SH$_2$$^+$ + C &  $ 0.4 $ &  $1.70\times 10^{-9}$ & 3.32 &  1 \\
&& $\rightarrow$ H$_2$CS + CH$_3$$^+$ & $ 0.5 $ &  $1.70\times 10^{-9}$ & 3.32 &  1 \\
\hline
19 & CH$_3$SH + O$_2$$^+$ & $\rightarrow$ CH$_3$SH$^+$ + O$_2$ & $ 0.5 $ &  $1.23\times 10^{-9}$ & 3.32 &  1 \\
\hline
20 & CH$_3$SH + H$^+$ & $\rightarrow$ CH$_3$SH$^+$ + H & $ 0.5 $ &  $5.5\times 10^{-9}$ & 3.32 & 1  \\
& & $\rightarrow$ CH$_3$$^+$ + H$_2$S  &  $ 0.25  $ & $5.5\times10^{-9}$ & 2.3 & 1 \\
& & $\rightarrow$ HCS$^+$ + H$_2$ + H$_2$  & $ 0.25  $ & $5.5\times10^{-9}$ & 2.3  & 1 \\
\hline
21 & CH$_3$SH + Photon & $\rightarrow$ H$_2$CS + H$_2$ & $1.40\times 10^{-9}$ & 0 & 2.28  & 9 \\
 & & $\rightarrow$     CH$_3$SH$^+$ + e$^-$ & $7.00\times 10^{-10}$ & 0 & 2.57  & 9 \\
& & $\rightarrow$ HS + CH$_3$ &  $1.20\times 10^{-9}$ & 0 & 2.28 & 9 \\
\hline
22 & CH$_3$SH + CRP & $\rightarrow$ H$_2$CS + H$_2$ & $3.17\times 10^{3}$ & 0 & 0 & 9 \\
 & & $\rightarrow$     CH$_3$SH$^+$ + e$^-$ &  $1.44\times 10^{3}$ & 0 & 0 &  9\\
& & $\rightarrow$ HS + CH$_3$ & $1.50\times 10^{3}$ & 0 & 0 & 9 \\
\hline
23 & CH$_3$SH$^+$ + e$^-$ & $\rightarrow$ HS + CH$_3$ & $3.00\times 10^{-7}$ & -0.5 & 0 & 9 \\
& & $\rightarrow$ H$_2$ + H$_2$CS& $3.00\times 10^{-7}$ & -0.5 & 0 & 9 \\
\hline
24 & CH$_3$SH$_2$$^+$ + e$^-$ & $\rightarrow$ H$_2$S + CH$_3$ &  $8.00\times 10^{-8}$ & -0.59 & 0 & 8 \\
& & $\rightarrow$ H + H$_2$CS + H$_2$ & $5.90\times 10^{-8}$ & -0.59 & 0 & 8 \\
& & $\rightarrow$ H + CH$_3$ + HS & $4.50\times 10^{-7}$ & -0.59 & 0 & 8 \\
& & $\rightarrow$ H + CH$_2$ + H$_2$S & $1.90\times 10^{-7}$ & -0.59 & 0  & 8\\
& & $\rightarrow$ H + CH$_3$SH &  $2.70\times 10^{-8}$ & -0.59 & 0 & 8 \\
& & $\rightarrow$ H$_2$ + CH$_3$S & $5.30\times 10^{-8}$ & -0.59 & 0 & 8  \\
\hline
1 & s-H + s-CS & $\rightarrow$ s-HCS & $ 1 $ &  0 & 0 & 9 \\
\hline
2 & s-H + s-CH$_3$S & $\rightarrow$ s-CH$_3$SH & $ 1 $ &  0 & 0 & 9 \\
\hline
3 & s-H + s-H$_2$CS & $\rightarrow$ s-CH$_3$S & $ 1 $ &  0 & 0 & 9  \\
\hline
4 & s-S + s-CH$_3$ & $\rightarrow$ s-CH$_3$S & $ 1 $ &  0 & 0 & 9 \\
\hline
\end{tabular}
\end{center}}
1 Rate constant calculated considering capture theory (using $\mu=1.56$ Debye and  $\alpha = 5.4$ Angstrom). Good agreement with \citep{1993JPCRD..22.1469A}, branching ratio assumed from \citep{1993JPCRD..22.1469A} . \\
2 Following H + CH$_3$O reaction from \citep{Hoyermann1981831,FT9918702331}   \\
3 Considering the various C reactions.   \\
4 Following  C + CH$_3$OH reaction from \citep{C4RA03036B}\\
5 Following O + CH$_3$O reaction from \citep{BBPC:BBPC19870910705} \\
6  Following \citep{doi:10.1021/j100297a042,doi:10.1021/jp9914828} \\
7 Following \citep{doi:10.1021/jp0108061}, branching ratio guessed from the OH + CH$_3$O reaction\\
8 Following \citep{2006FaDi..133..177G}\\
9 Following the similar reactions of CH$_3$OH and its related species\\
10 Following \citep{1982Apss..87..1}  \\
\end{table*}



\bsp	
\label{lastpage}
\end{document}